\documentclass[english,twoside,a4paper,10pt]{article}

\usepackage[latin1]{inputenc}
\usepackage[T1]{fontenc}
\usepackage{amsmath}
\usepackage{amsfonts}
\usepackage{graphicx}
\usepackage{subfigure}
\usepackage{a4wide}
\usepackage{amssymb}
\usepackage{fancyhdr}
\usepackage{mathrsfs}
\usepackage[toc,page]{appendix}

\begin{document}

\title{\bf Warm inflation in Horndeski gravity}
\author{ 
Lorenzo Sebastiani$^{1,2}$\footnote{E-mail: lorenzo.sebastiani@unitn.it
},\,\,\,
Shynaray Myrzakul$^{3}$\footnote{E-mail: shynaray1981@gmail.com},\,\,\,
Ratbay Myrzakulov$^{1}$\footnote{E-mail: rmyrzakulov@gmail.com}\\
\\
\begin{small}
$^{1}$ Department of General \& Theoretical Physics and Eurasian Center for
\end{small}\\
\begin{small} 
Theoretical Physics, Eurasian National University, Astana 010008, Kazakhstan
\end{small}\\
\begin{small}
$^2$ Dipartimento di Fisica, Universit\`a di Trento, Italy 
\end{small}\\
\begin{small}
$^3$ Department of Theoretical and Nuclear Physics,
\end{small}\\
\begin{small}
Al-Farabi Kazakh National University, Al-Farabi Almaty, Kazakhstan
\end{small}
}

\date{}

\maketitle


\begin{abstract}
In this paper, we investigate a class of Horndeski scalar-tensor theory of gravity for warm inflation. We present some models where the early-time acceleration is realized in the weak and in the strong dissipation regime. Cosmological perturbations are analyzed.
\end{abstract}



\tableofcontents
\section{Introduction}

Since its first proposal by Guth~\cite{Guth} and Sato~\cite{Sato} in 1981, the inflationary paradigm is at the basis of the theory for the early-time evolution of our Universe (see Refs.~\cite{Linde, revinflazione} for some reviews).  The success of such a theory is due to the fact that, if the Universe underwent a period of strong accelerated expansion after the Big Bang, we can solve the problem of initial conditions of our Friedmann Universe and obtain the description of its inhomogeneities at the galactic scale.  

Usually, the early-time inflation is driven by a scalar field, dubbed ``inflaton'', subjected to some suitable potential. In the standard scenario, since during inflation any field of radiation or ultrarelativistic matter is shifted away, at the end of inflation some reheating mechanism to convert the energy of inflaton in particle fields is required. However, if one introduces a coupling between inflaton and radiation, the energy density of radiation can be mantained almost a constant during inflation and the reheating is no longer necessary. In this sense, due to its physical implications, the warm inflationary scenario deserves some special attention 
~\cite{Berera00, Berera0, Berera01, Berera1, Berera2, warm4, warm5, warm6, warm7, warm8}. 

Warm inflation has been recently studied in the framework of many different theories, for instance in multidimensional braneworld models~\cite{Campo, miobrane}.  
We note that at high energy some curvature corrections, maybe motivated by quantum effects, may emerge from the theory modifiying Einstein's gravity~\cite{R1, R2, R3, R4,  RGinfl, Odinfrev}. In this respect, one interesting class of modifying theories of gravity is represented by the scalar-tensor models of Horndeski gravity~\cite{Horn}, where, despite the involving form of the Lagrangian, the equations of motion appear at the second order like in General Relativity. In the last years many works about Horndeski gravity have been carried out, especially related to the early-time inflation~\cite{Amendola, Def, DeTsu, DeFelice, Kob, Kob2, Qiu, EugeniaH, mioH, cognoH}. 

In this work, we would like to investigate a class of Horndeski scalar-theory where the Horndeski field is coupled with the Einstein's tensor in the context of the warm inflationary scenario. We will achieve our results by introducing a coupling between inflaton and radiation, in order to recover the radiation dominated Universe at the end of inflation. We will show how it is possible to obtain realistic scenarios for the early-time acceleration in the both strong and weak dissipation regimes. Cosmological perturbations will also be analyzed and confronted with the last Planck satellite data~\cite{Planck}. 

The paper is organized in the following way. In Section {\bf 2} we will present our Horndeski model and its Friedmann equations for warm inflation. Here, the slow-roll approximation will be adopted to describe inflation. Furthermore, two separate subsections will be devoted to the analysis of some inflationary models in the weak dissipation regime and in the strong dissipation regime, by assuming different forms of coupling between inflaton and radiation. In Section {\bf 3} we will present a reconstruction tecnique which permits to find the models by starting from a specific form of inflationary solution. In Section {\bf 4} we will discuss the cosmological perturbations and the spectral index and the tensor-to-scalar ration will be derived. Conclusions and final remarks are given in Section {\bf 5}.

We use units of $k_{\mathrm{B}} = c = \hbar = 1$ and 
$8\pi/M_{Pl}^2=1$, where $M_{Pl}$ is the Planck Mass.

\section{The model}

In this paper, we will consider a very popular subclass of Horndeski scalar-tensor theory~\cite{Horn}, where the scalar field is non-minimally coupled with gravity through the Einstein's tensor. As a result, one deals with higher curvature corrections to Einstein's gravity at the large curvature, when inflation takes place. Moreover, in order to get the warm scenario, we will introduce an interaction between the scalar field (i.e. the inflaton) and the radiation field. The action of the model reads,
\begin{equation}
I=\int_\mathcal{M}d^4x\sqrt{-g}\left[\frac{R}{2}-\frac{g_{\mu\nu}\partial^\mu\phi\partial^\nu\phi}{2}-V(\phi)+\alpha G_{\mu\nu}\partial^\mu\phi\partial^\nu\phi+\mathcal L_r+\mathcal L_{int}\right]\,,\label{action}
\end{equation}
where $\mathcal M$ is the space-time manifold, $g$ is the determinant of the metric tensor $g_{\mu\nu}$, $R$ is the Ricci scalar of the Hilbert-Einstein term of General Relativity (GR), $V(\phi)$ is a potential of the scalar field $\phi$, $\alpha$ is a constant parameter and $G_{\mu\nu}\equiv \left[R_{\mu\nu}-g_{\mu\nu}R/2\right]$ is the usual Einstein's tensor, $R_{\mu\nu}$ being the Ricci tensor. Finally, we have added the Lagrangian $\mathcal L_r$ of the radiation contents of the Universe, while $\mathcal L_{int}$ describes the interaction between inflaton and radiation.

Let us consider the flat Friedmann-Robertson-Walker (FRW) metric, 
\begin{equation}
ds^2=-dt^2+a(t)^2 d{\bf x}^2\,,\label{metric}
\end{equation}
with $a(t)$ the scale factor of the Universe. An useful parameterization often used in describing the dynamical evolution of the system is obtained by making use of the $e$-folds number,
\begin{equation}
N=\log\left[\frac{a(t_0)}{a(t)}\right]\,,\label{N}
\end{equation}
where $a(t_0)$ is the scale factor at the fixed time $t_0$. In the specific, we will set 
$t_0$ as the time when inflation ends, such that $0< N$ (i.e. $t<t_0$) during inflation. 

The variation of the action (\ref{action}) with respect to the metric gives~\cite{mioH},
\begin{equation}
3H^2 -9\alpha H^4\phi'^2=
\frac{H^2\phi'^2}{2}+V(\phi)+\rho_{r}
\,,\label{EOM1}
\end{equation}
\begin{eqnarray}
\left(2H H'-3H^2\right)&=&\frac{H^2\phi'^2}{2}-V(\phi)+\alpha(H^2\phi'^2)
\left(3H^2-2H H'\right)-6\alpha H^4 \phi'^2
\nonumber\\
&&
+4\alpha H^2\phi'\left(H^2\phi''+H H'\phi'\right)+4\alpha H^3 H' \phi'^2\,,
\end{eqnarray}
where $\rho_r$ is the energy density of radiation and the prime denotes the derivative with respect to the $e$-folds number $N$.

The derivative with respect to the field together with the conservation law of radiation read,
\begin{equation}
H^2\phi'\phi''(1+6\alpha H^2)+\phi'^2 H' H(1+18\alpha H^2)
-3\phi'^2 H^2(1+6\alpha H^2)+V_\phi(\phi)\phi'=\mathcal Y f(\phi) H\phi'^2\,,\label{conslawphi}
\end{equation}
\begin{equation}
\rho_r'-4\rho_r=-\mathcal Y f(\phi) H\phi'^2\,,\quad 0<\mathcal Y, f(\phi) \,,\label{conslawrad}
\end{equation}
where we assume that the interaction between the scalar field and radiation leads to the friction term $-\mathcal Y f(\phi) H \phi'$, $\mathcal Y$ being a positive constant and $f(\phi)$ a function of the field~\cite{Berera1, Berera2}. We observe that its contribute is significative for the dynamic of the scalar field as long as $H\leq \mathcal Yf(\phi)$, but even when $ \mathcal Yf(\phi)\leq H$ it plays an important role in the continuity equation of radiation. We note that the positivity of $\mathcal Y f(\phi)$ is required to get a positive energy density of radiation during inflation.

In order to describe inflation we must introduce  the $\epsilon$ slow-roll parameter,
\begin{equation}
\epsilon=\frac{H'}{H}\,,\label{epsilon}
\end{equation}
which has to be small and positive in order to obtain a strong accelerated expansion. Therefore, the early-time acceleration ends when $\epsilon$ is on the order of the unit. 

At the time of inflation the Hubble parameter is large and almost a constant. Thus, the (quasi) de Sitter expansion takes place under the slow-roll approximation with $\phi'^2\ll 1$ and $|\phi''|\ll |\phi'|$, such that Eqs. (\ref{EOM1}) and (\ref{conslawphi}) read
\begin{equation}
3H^2 -9\alpha H^4\phi'^2\simeq
V(\phi)
\,,\label{EOM1bis}
\end{equation}
\begin{equation}
-3\phi' H^2(1+6\alpha H^2)+V_\phi(\phi)\simeq\mathcal Y f(\phi) H\phi'\,,\label{conslawphibis}
\end{equation}
where we have taken into account that the contribution of the field is dominant in the balance of energy of the inflationary Universe and $\epsilon\ll 1$. In order to combine the effects from the higher curvature corrections of GR with the canonical scalar field description, we assume $|\alpha|\sim 1/H^2$ during inflation, such that we get,
\begin{equation}
3H^2\simeq V\,,\quad
\phi'^2\simeq\frac{V'}{3H^2(1+6\alpha H^2)\left(1+r\right)}\,,\quad r=\frac{\mathcal Y f(\phi)}{3H(1+6\alpha H^2)}\,,\label{set1}
\end{equation}
where $V\equiv V(N)$ and we have introduced the adimensional parameter $r$.  
The strong and the weak dissipation regimes correspond to $1\ll r$ and  $r\ll  1$, respectively.
We see that the de Sitter expansion is well described by the potential of the field like in the classical inflationary scenario from scalar field models, while the Horndeski corrections and the interaction between inflaton and radiation modify the exit from inflation. We should note that, even if in the strong dissipation regime $1\ll r$ the Horndeski corrections drop down in the second equation of (\ref{set1}), they may play a role in the perturbation theory.

Finally, one has from (\ref{conslawrad}),
\begin{equation}
\rho_{r}\simeq \frac{\mathcal Y f(\phi) H\phi'^2}{4}=\frac{3H^2(1+6\alpha H^2)r\phi'^2}{4}\,,\label{set2}
\end{equation}
and the energy density of radiation remains almost a constant. If one looks for the ratio between the energy density of radiation and the energy density of the scalar field $\rho_\phi= \left[H^2\phi'^2/2+V(\phi)\right]$ we obtain,
\begin{equation}
\frac{\rho_r}{\rho_\phi}\simeq\frac{r\epsilon}{2(1+6\alpha H^2)\left(1+r\right)}\,.
\label{ratio}
\end{equation}
It means that during inflation, when $\epsilon\ll 1$, the energy density of radiation is negligible, but when the early-time acceleration ends and $\epsilon\simeq 1$, radiation emerges in the cosmological scenario without invoking any reheating.

In what follows we will propose some explicit examples of Horndeski models for warm inflation. For the field potential we will take a power-law form as
\begin{equation}
V(\phi)=\frac{\lambda(-\phi)^n}{n}\,,\quad 0<\lambda\,,n\,,\label{V0}
\end{equation}
where $\lambda\,, n$ are positive parameters. The sign minus in front of the field has been introduced since we require that during inflation the field is negative and its magnitude very large, while goes to zero at the end of inflation.
The direct consequence of (\ref{V0}) is that the solution of the Hubble parameter during the early-time accelerated expansion is given by,
\begin{equation}
H^2\simeq \frac{\lambda(-\phi)^n}{3n}\,.\label{HDS}
\end{equation}
By starting from this solution, we will investigate different kinds of models where the coupling between inflaton and radiation brings to a strong or a weak dissipation regimes.

\subsection{Strong dissipation regime}

The strong dissipation regime corresponds to $1\ll r$ in (\ref{set1}), namely
\begin{equation}
3H(1+6\alpha H^2)\ll \mathcal Y f(\phi)\,.
\end{equation}
Thus, by taking into account (\ref{HDS}), we may choose
\begin{equation}
f(\phi)=f_0(-\phi)^{m/2}\,,\quad n<m\,,\label{fs}
\end{equation}
where $f_0$ is a positive constant and $m$ a (positive) number larger than $n$. In these expressions, we are considering $\alpha$ on the order of $|\alpha|\sim 1/H^2$. Thus, from the second equation in (\ref{set1}) with $V'=V_\phi(\phi)/\phi'$ and $1\ll r$ one has
\begin{equation}
\phi(N)=-\phi_0 N^{\frac{2}{4+m-n}}\,,
\end{equation}
where
\begin{equation}
\phi_0=\left(\frac{(4+m-n)\sqrt{3n\lambda}}{2\mathcal Y f_0}\right)^{\frac{2}{4+m-n}}\,.
\end{equation}
We see that, thanks to the fact that $n<m$, the magnitude of the field at the beginning of inflation when $1\ll N$ is large and leads to a large Hubble parameter, while $\phi'^2\propto 1/N^{(2+m-n)/(4+m-n)}$ and tends to vanish.

The Hubble parameter and the $\epsilon$ slow-roll parameter (\ref{epsilon}) behave as
\begin{equation}
H^2=\frac{\lambda}{3n}\phi_0^n N^{\frac{2n}{4+m-n}}\,,\quad\epsilon=\frac{n}{(4+m-n) N}\,.
\end{equation}
We can now evaluate the amount of radiation at the time of the early-time accelerated expansion. In the strong dissipation regime $1\ll r$ one gets from (\ref{ratio}),
\begin{equation}
\frac{\rho_r}{\rho_\phi}\simeq\frac{n}{(4+m-n) N}\left(\frac{1}{2(1+6\alpha H^2)}\right)\,,
\end{equation}
and the radiation is negligible when $1\ll N$. On the other hand, for small values of $N$, the energy density of radiation becomes dominant and we enter in the weak dissipation regime with $r\ll 1$. At this point, the friction term between inflaton and radiation becomes negligible, and the radiation dominated expansion takes place. 

In the next subsection, we will analyze the case of weak dissipation regime during the accelerated expansion.

\subsection{Weak dissipation regime}

The weak dissipation regime corresponds to $r\ll 1$ in (\ref{set1}), namely
\begin{equation}
\mathcal Y f(\phi)\ll 3H(1+6\alpha H^2)\,.
\end{equation}
Thus, given the potential (\ref{V0}) with solution (\ref{HDS}), a reasonable choice may be
\begin{equation}
f(\phi)=f_0(-\phi)^{m/2}\,,\quad m<n\,,\label{fr}
\end{equation}
with $f_0$ a positive parameter and $m$ a (positive) number smaller than $n$. Now the Horndeski correction plays an important role in the second equation of (\ref{set1}), and in order to get an explicit solution we will fix the parameter $n$ as
\begin{equation}
n=2\,.
\end{equation}
In this case, in the limit $r\ll 1$, the second equation in (\ref{set1}) leads to
\begin{equation}
\phi(N)=-\frac{1}{\sqrt{\alpha\lambda}}\sqrt{\sqrt{8\alpha\lambda N+1}-1}\,.\label{fr1}
\end{equation}
We can see again that the magnitude of $\phi$ is large and quasi a constant ($\phi'^2\ll 1$) when $1\ll N$ and tends to vanish at the end of inflation when $N\ll 1$. 

We can express the Hubble parameter and the $\epsilon$ slow-roll parameter in terms of the $e$-folds number as,
\begin{equation}
H^2=\frac{\sqrt{1+8\alpha\lambda N}-1}{6\alpha}\,,\quad 
\epsilon\simeq \frac{1}{4N}\,,\label{fr2}
\end{equation}
and inflation takes place at $1\ll N$. Moreover, during the accelerated expansion, the ratio between the energy density of radiation and inflaton reads,
\begin{equation}
\frac{\rho_r}{\rho_\phi}\simeq\frac{2^{(3m-26)/8}f_0\mathcal Y}{\sqrt{3}N^{(2-m)/8}\lambda^{(2+m)/8}\alpha^{(m-2)/8}(1+2\sqrt{2\alpha\lambda N})}\,,
\end{equation}
and remains small during the accelerated phase with $1\ll N$, while grows up at the end of inflation when $N\ll 1$.

\section{Reconstruction of Horndeski models for warm inflation}

In order to study some different evolutions of our Horndeski model during the inflationary phase, we can also start by fixing some specific behaviour of the Hubble parameter. In this section, we will consider one example that does not belong to the class of solutions investigated above.

Let us take,
\begin{equation}
H^2=H_0^2\left[1-\frac{1}{(N+1)^\lambda}\right]\,,\quad 1\leq\lambda\,,
\label{lastex}
\end{equation}
where $H_0$ is a (positive) constant and $\lambda$ a number larger or equal to one. This solution tends to the de Sitter space-time for large values of $N$, while vanishes at $N=0$. When $1\ll N$ one has,
\begin{equation}
\epsilon=\frac{\lambda}{2(N+1)^{\lambda+1}}\,.\label{29}
\end{equation}
By using the first equation in (\ref{set1}) we derive the on-shell form of the potential,
\begin{equation}
V(N)=3H_0^2\left[1-\frac{1}{(N+1)^\lambda}\right]\,.
\end{equation}
In order to carry out our investigation, we will distingue two cases, namely the strong dissipation regime and the weak dissipation regime. 

For the strong dissipation regime, we may consider a friction term in the form,
\begin{equation}
f(N)=f_0\left[(N+1)^\zeta-1\right]\,,\quad 0<\zeta\,,
\end{equation}
where $f_0$ is a positive constant and $\zeta$ a positive number. This kind of function has been constructed in order to make vanishing the coupling between inflaton and radiation at the end of inflation when $N=0$. Thus, in the limit $1\ll N$, the second equation in (\ref{set1}) leads to,
\begin{equation}
\phi'(N)^2\simeq\frac{3H_0\lambda}{f_0\mathcal Y(1+N)^{1+\lambda+\zeta}}\,.\label{32}
\end{equation}
Finally, the ratio between the energy density of inflaton and radiation in (\ref{ratio}) reads,
\begin{equation}
\frac{\rho_r}{\rho_\phi}\simeq\frac{\lambda}{4(1+6H_0^2\alpha)(1+N)^{1+\lambda}}\,,\label{33}
\end{equation}
and is small for $1\ll N$, while grows up when $N$ tends to zero. 

Finally, we can reconstruct our model by giving the explicit expressions for the potential and the friction term. From (\ref{32}) we have
\begin{equation}
\phi=-\phi_0\left[1-(1+N)^{\frac{(1-\zeta-\lambda)}{2}}\right]\,,\quad (N+1)=\left(\frac{\phi_0}{\phi+\phi_0}\right)^{\frac{2}{\zeta+\lambda-1}}\,,
\end{equation}
where the magnitude of the field at the beginning of inflation is given by,
\begin{equation}
\phi_0=\frac{2}{\lambda+\zeta-1}\sqrt{\frac{3H_0\lambda}{f_0\mathcal Y}}\,.
\end{equation}
As a consequence, one obtains
\begin{equation}
V(\phi)=3H_0^2\left[1-\left(\frac{\phi+\phi_0}{\phi_0}\right)^{\frac{2\lambda}{\zeta+\lambda-1}}\right]\,,\quad
f(\phi)=f_0\left[\left(\frac{\phi_0}{\phi+\phi_0}\right)^{\frac{2\zeta}{\zeta+\lambda-1}}-1\right]\,.\label{36}
\end{equation}
In order to reproduce the solution (\ref{lastex}) in the weak dissipation regime, we can use a friction term in the form,
\begin{equation}
f(N)=f_0\frac{1}{(N+1)^\zeta}\,,\quad 0<\zeta\,,
\end{equation}
with $f_0$ a positive constant and $\zeta$ a positive number again. In this case the second equation in (\ref{set1}) leads to
\begin{equation}
\phi'^2(N)\simeq\frac{\lambda}{(1+6\alpha H_0^2)(1+N)^{1+\lambda}}\,,\label{37}
\end{equation}
while the ratio between the energy density of radiation and inflaton can be computed as
\begin{equation}
\frac{\phi_r}{\phi_\phi}\simeq\frac{\mathcal Y f_0\lambda}{12H_0(1+6\alpha H_0^2)^2(1+N)^{1+\lambda+\zeta}}\,,
\label{39}
\end{equation}
and is small when $1\ll N$. 

Now from (\ref{37}) we get,
\begin{equation}
\phi(N)=-\phi_0\log\left[
1+N
\right]\,,\quad (N+1)=\text{e}^{-\phi/\phi_0}\,,\quad \lambda=1\,,
\end{equation}
\begin{equation}
\phi(N)=-\phi_0\left(1-(N+1)^\frac{(1-\lambda)}{2}\right)\,,\quad
(N+1)=\left(\frac{\phi_0}{\phi+\phi_0}\right)^{\frac{2}{\lambda-1}}\,,\quad 1<\lambda\,,
\end{equation}
with
\begin{equation}
\phi_0=\frac{1}{\sqrt{1+6\alpha H_0^2}}\,,\quad \lambda=1\,,
\end{equation}
\begin{equation}
\phi_0=\frac{2}{\lambda-1}\sqrt{\frac{\lambda}{1+6\alpha H_0^2}}\,,\quad 1<\lambda\,.
\end{equation}
Thus, the model is fully reconstructed as
\begin{equation}
V(\phi)=3H_0^2\left(1-\text{e}^{\phi/\phi_0}\right)\,,\quad f(\phi)=f_0\text{e}^{\zeta\phi/\phi_0}\,,\quad \lambda=1\,,\label{420}
\end{equation}
\begin{equation}
V(\phi)=3H_0^2\left[1-\left(\frac{\phi+\phi_0}{\phi_0}\right)^{\frac{2\lambda}{\lambda-1}}\right]\,,\quad
f(\phi)=f_0\left(
\frac{\phi+\phi_0}{\phi_0}
\right)^\frac{2\zeta}{\lambda-1}\,,\quad 1<\lambda\,,\label{42}
\end{equation}
and we see that the potential has the same form of (\ref{36}) without the contribution from the coupling between inflaton and radiation ($\zeta\rightarrow 0$).\\
\\
At the end of inflation the Hubble parameter in (\ref{lastex}) tends to vanish and the higher curvature Horndeski corrections 
in the gravitational action do not play longer any significative role. Moreover,
the 
energy density of radiation becomes dominant with respect to the energy density of inflaton, as it is showed in (\ref{33}) and in (\ref{39}) for small values of $N$. In other words, we have a phase transition and the constant (vacuum) energy of radiation survives to the early-time acceleration.
Thus, one obtains a radiation dominated Universe without invoking any prereheating or reheating mechanism and the Friedmann expansion takes place like in General Relativity.\\ 
\\
We have seen how several different Horndeski models can reproduce the early-time acceleration in the context of warm inflation. However, the scenario becomes realistic only if it is possible to recover the correct spectrum of the inhomogeneities at the galactic scale in our observable Universe. In the next section we will analyze this point.

\section{Cosmological perturbations}

The scalar perturbations enter in the perturbed FRW metric as,
\begin{equation}
ds^2=-(1+2\zeta(t, {\bf x}))dt^2+a(t)^2(1-2\psi(t, {\bf x}))d{\bf x}\,,\label{FRWp}
\end{equation}
where we have used the Newton's gauge and $|\zeta|\equiv |\zeta(t, {\bf x})|\ll 1$ and $|\psi|\equiv |\psi(t, {t,\bf x})|\ll 1$ are functions of the space-time coordinates. The field equations immediately contraint these quantities as $\psi=\zeta$. We must now distinguish between the cases of inflaton in weak dissipation regime and inflaton in strong dissipation regime.\\  
\\
In the weak dissipation regime with $r\ll 1$, the perturbations of the radiation field are negligible and the oscillations of the inflaton field are generated by quantum fluctuations. As a consequence,
in order to derive the cosmological scalar perturbations inside our Horndeski model we can use the formalism well developed in Refs.~\cite{Def, DeTsu, DeFelice}, with the account of the coupling between inflaton and radiation.

The equation for perturbations assumes the form,
\begin{equation}
\ddot v-\frac{c_s^2}{a^2}\bigtriangleup v-\frac{\ddot z}{z}v=-c_s^2 \frac{\dot\phi^2\mathcal Y f(\phi)}{6H}v\,,\quad 
v=z\zeta\,,\quad z=\sqrt{a^3 Q}\,,
\end{equation}
where the dot denotes the time derivative and $Q\,,c_s^2$ are functions of the field and the Hubble parameter evaluated on the quasi de Sitter solution of the background.  They are given by:
\begin{equation}
Q=\frac{(1+6\alpha H^2)(1+r)\dot\phi^2}{2H^2}\,,\label{Q0}
\end{equation}
\begin{equation}
c_s^2=-\frac{2H\dot H}{(H+6\alpha H^3)(1+r)\dot\phi^2}\,,\label{c0}
\end{equation}
where we have considered $|\alpha|\sim 1/H^2$ and $\phi'^2\ll 1$. 
When $\mathcal Y=0$ we recover the results in Refs.~\cite{Def, DeTsu, DeFelice}, while the friction term has been introduced by following the line of Ref.~\cite{miobrane}. 

Now one can decompose 
$v$ in Fourier modes $v=v_k(t)\exp[i {\bf k}{\bf x}]$, such that they are governed by the following equation,
\begin{equation}
\ddot v_k(t)+\left(k^2\frac{c_s^2}{a^2}-\frac{\ddot z}{z}\right)v_k(t)=-c_s^2 \frac{\dot\phi^2\mathcal Y f(\phi)}{6H} v_k(t)\,.\label{eqpert}
\end{equation}
The cosmological perturbations can propagate only if the speed of sound $c_s^2$ is different to zero, otherwise the spectral index will be flat. 
In the specific, in terms of the e-folds (\ref{N}), the speed of sound is derived as  
\begin{equation}
c_s^2\simeq\frac{2H'}{(H+6H^3\alpha)(1+r)\phi'^2}=1\,,\label{sound}
\end{equation}
which is the classical result for canonical scalar field inflation.

The solution of Eq.~(\ref{eqpert}) for long-wave perturbations with $k^2/a^2\ll 1$ in the quasi de-Sitter space-time is given by,
\begin{equation}
v_k(t)\simeq c_0\sqrt{\frac{a}{2}}\frac{a H}{(c_s k)^{3/2}}
\text{e}^{\pm i k\int \frac{c_s}{a}dt}
\left(
\text{e}^{\Gamma(t)}
\right)
\left(1+i c_s k\int\frac{dt}{a}\right)\,,\label{res00}
\end{equation}
where the function $\Gamma(t)$ encodes the contribution from the interaction beteween the inflaton and the radiation field and reads,
\begin{equation}
\Gamma(t)=-\frac{c_s^2}{18}\int
\frac{\dot\phi^2\mathcal Y f(\phi)}{H^2} dt\,. 
\end{equation}
Finally, the constant $c_0$ in (\ref{res00}) must be fixed by the  Bunch-Davies vacuum state\\ 
$v_k(t)=\sqrt{a}\exp[\pm i \kappa\int c_s dt/a]/(2\sqrt{c_s\kappa})$ in the asymptotic past, namely $c_0=i/\sqrt{2}$.
One obtains,
\begin{equation}
\zeta_k\equiv \frac{v_k(t)}{\sqrt{Q a^3}}=i\frac{H}{2\sqrt{Q}(c_s k)^{3/2}}
\left(
\text{e}^{\Gamma(t)}
\right)\text{e}^{\pm i k\int \frac{c_s}{a}dt}
\left(1+i c_s k\int\frac{dt}{a}\right)
\,.
\end{equation}
The variance of the power spectrum of perturbations must be evaluated on the sound horizon crossing $c_s\kappa\simeq H a$, namely
\begin{equation}
\mathcal P_{\mathcal R}\equiv\frac{|\zeta_k|^2 k^3}{2\pi^2}|_{c_s k\simeq H a}=\frac{H^2 \left(
\text{e}^{\Gamma(t)}
\right)}{8\pi^2 c_s^3 Q}|_{c_s k\simeq H a}\,.\label{55}
\end{equation}
As a consequence, by taking into account (\ref{Q0}) and by posing $c_s^2=1$, in terms of the $e$-folds the spectral index $n_s$ defined by
\begin{equation}
(n_s-1)=\frac{d\ln \mathcal P_{\mathcal R}}{d \ln k}|_{k=a H/c_s}\,,\label{56}
\end{equation}
results to be
\begin{equation}
(n_s-1)\simeq\frac{(3HH''\phi'-2HH'\phi'')(1+6\alpha H^2)-H'^2\phi'(7+54\alpha H^2)}{2H H'\phi'(1+6\alpha H^2)}-\frac{\phi'^2\mathcal Y f(\phi)}{18H}\,.
\end{equation}
We should observe that, generally, when warm inflation is realized in the weak dissipation regime, the friction term does not play a significative role in the perturbed theory. For example, in the case of model (\ref{V0}) with solution (\ref{HDS}) with $n=2$ and the friction term as in (\ref{fr}) such that $m<2$, the inflation described by (\ref{fr1})--(\ref{fr2}) leads to a spectral index $(n_s-1)\simeq -3/(2N)$. Furthermore, the model in (\ref{420})--(\ref{42}) leads to $(n_s-1)\simeq- (1+\lambda)/N$. Considering that at the beginning of inflation $N\simeq 60$ and that the cosmological data~\cite{Planck} lead to $n_{\mathrm{s}} = 0.968 \pm 0.006$, we can conclude that these models can be viable, the second one for $\lambda=1$ as in (\ref{420}).\\
\\
In the strong dissipation regime with $1\ll r$ the oscillations of the inflaton field are no longer
generated by quantum fluctuations, but are given by thermal interactions
with the radiation field whose energy density follows the Stefan-
Boltzmann law,
\begin{equation}
\rho_r=\beta T^4\,,\label{Bolz}
\end{equation}
where $T$ is the temperature of the thermal bath of inflaton and radiation and $\beta$ is a positive constant. In our analysis we will follow Ref.~\cite{warm8}, where Galileon $G$-warm inflationary models have been considered. By perturbing the field as $\phi\rightarrow\phi+\delta\phi(t, {\bf x})$ and by decomposing $\delta\phi(t, {\bf x})$ in Fourier modes $\delta(t, {\bf x})=\delta\phi_k(t)\text{e}^{-i{\bf k}{\bf x}}$,
from the
second equation in (\ref{set1}) one has,
\begin{equation}
\dot\delta_k\phi(t)\simeq\frac{1}{3H(1+r)}\left(-(k^2+V_{\phi\phi})\delta\phi_k(t)+\xi({\bf{k}},t)\right)\,,
\label{sd1}
\end{equation}
where $\xi({\bf{k}},t)$ is a thermal stocastic noise such that $<\xi(t)>=0$. The solution of (\ref{sd1}) reads,
\begin{equation}
\delta\phi_k(t)\simeq\theta(t)\int\frac{\xi(t)}{3H(1+r)\theta(t)}dt\,,
\end{equation}
where
\begin{equation}
\theta(t)=\exp
\left[
-\int \left(\frac{k^2+V_{\phi\phi}}{3H(1+r)}\right)dt
\right]\,.\label{theta}
\end{equation}
The thermal fluctuation of the scalar field is given by~\cite{Berera0, Berera1, Berera2},
\begin{equation}
\delta\phi^2=\frac{k_F T}{2\pi^2}\,,
\end{equation}
where $k_F$ corresponds to the freeze out wave number and follows from (\ref{theta}) as
\begin{equation}
\left(\frac{k_F^2}{3H^2(1+r)}\right)=1\,.
\end{equation}
Here, we considered $|V_{\phi\phi}|\ll k^2$ and we posed the time interval $\Delta t\simeq 1/H$. 
The temperature is derived from (\ref{set2}) and (\ref{Bolz}) as,
\begin{equation}
T^4=\frac{3(1+6\alpha H^2)r\dot\phi^2}{4\beta}\,.
\label{64}
\end{equation}
Finally, the power spectrum of the scalar perturbations follows from the relation~\cite{Berera0, Berera1},
\begin{equation}
\mathcal P_{\mathcal R}=\left(\frac{H}{\dot\phi}\right)^2\delta\phi^2\,.
\end{equation}
Thus, we get,
\begin{equation}
\mathcal P_{\mathcal R}=\frac{1}{2\pi^2}\left(\frac{H}{\dot\phi}\right)^2
\left(
\frac{\mathcal Y f(\phi)\dot\phi^2}{4\beta H}
\right)^{1/4}\sqrt{3H^2(1+r)}\,.
\label{PPP}
\end{equation}
Now, by taking $c_s^2=1$, one has for the spectral index (\ref{56}),
\begin{equation}
(n_s-1)=\frac{9\dot H}{4H^2}-\frac{3\ddot\phi}{2H\dot\phi}+\frac{3\dot\phi f_\phi(\phi)}{4Hf(\phi)}-\frac{6\alpha \dot H}{1+6\alpha H^2}\,.
\end{equation}
For $\alpha=0$ we recover the result of Ref.~\cite{Berera2}. In terms of the $e$-folds number (\ref{N}) we obtain,
\begin{equation}
(n_s-1)=-\frac{3H'}{4H}+\frac{3\phi''}{2\phi'}-\frac{3 f'(\phi)}{4f(\phi)}+\frac{6\alpha H H'}{(1+6\alpha H^2)}\,.\label{n2}
\end{equation}
For example, in the case of the model (\ref{V0}) with solution (\ref{HDS}), when $f(\phi)$ is given by (\ref{fs}) with $n<m$, inflation is realized in the strong dissipation regime and Eq.~(\ref{n2}) leads to 
\begin{equation}
(n_s-1)\simeq\frac{-12-9m+7n}{4N(4+m-n)}\,.
\end{equation}
This result is in agreement with the Planck data for $N\simeq 60$ when $(n_s-1)\simeq -2/N$, namely
\begin{equation}
m=20-n\,.
\end{equation}
For the model (\ref{36}) with solution (\ref{lastex}) one has
\begin{equation}
(n_s-1)\simeq-\frac{3(1+2\zeta+\lambda)}{4N}\,,
\end{equation}
such that we must require 
\begin{equation}
\zeta=\frac{5-3\lambda}{6}\,.
\end{equation}
Since $\zeta$ has to remain a positive quantity, $\lambda$ must belong to the range $1\leq\lambda<5/3$.\\
\\
The amplitude of the tensorial perturbations return be the classical ones of Horndeski gravity, namely, 
\begin{equation}
\mathcal P_{\mathcal T}\simeq \frac{8}{(1+\alpha\dot\phi^2)}\left(\frac{H}{2\pi}\right)^2\simeq 
8\left(\frac{H}{2\pi}\right)^2
\,.
\end{equation}
The tensor-to-scalar ratio $r\equiv \mathcal P_{\mathcal T}/P_{\mathcal R}$ in the weak dissipation regime follows from (\ref{55}) with $c_s=1$ as
\begin{equation}
r\equiv\frac{\mathcal P_{\mathcal T}}{P_{\mathcal R}}\simeq\frac{8 (1+6\alpha H^2)(1+r)\dot\phi^2}{H^2}\frac{1}{\text{e}^{\Gamma(t)}}\,,
\end{equation}
or, in terms of the $e$-folds number,
\begin{equation}
r\simeq 8 (1+6\alpha H^2)(1+r)\phi'^2\frac{1}{\text{e}^{\Gamma(N)}}=16\epsilon\frac{1}{{\text{e}^{\Gamma(N)}}}\,,\quad \Gamma(N)=\frac{1}{18}\int H\phi'^2\mathcal Y f(N)dN\,,
\end{equation}
where we have introduced the $\epsilon$ slow-roll parameter (\ref{epsilon}). For example, the solution in (\ref{fr2}) with (\ref{fr}) leads to $r<4/N$, while the one in (\ref{lastex})--(\ref{29}) leads to $\epsilon\leq 8\lambda/N^\lambda$, $1\leq\lambda$. Since the last cosmological data~\cite{Planck} constraint the tensor-to-scalar ratio as $r<0.11\, (95\%\,\text{CL})$, we can argue that for $N\simeq 60$ the values $r<8/$ or $r\sim 1/N^\lambda$, $1<\lambda$ are acceptable and the mentioned solutions are viable.

For the strong dissipation regime with power scalar spectrum (\ref{PPP}) the tensor-to-scalar ratio results to be,
\begin{equation}
r\equiv\frac{\mathcal P_{\mathcal T}}{P_{\mathcal R}}\simeq
\frac{4\dot\phi^2}{(1+\alpha\dot\phi^2)}\left(\frac{4\beta H}{\mathcal Y f\dot\phi^2}\right)^{1/4}\frac{1}{\sqrt{3H^2(1+r)}}\,,
\end{equation}
or, in terms of the $e$-folds number,
\begin{equation}
r\simeq
\frac{4H^2\phi'^2}{(1+\alpha H^2\phi'^2)}\left(\frac{4\beta }{\mathcal Y f H\phi'^2}\right)^{1/4}\frac{1}{\sqrt{3H^2(1+r)}}\,.
\end{equation}
In this case we observe that, when $1\ll r$, this quantity will be easily suppressed, rendering the predictions of the models in agreement with the cosmological data.

\section{Conclusions}

In this paper, a popular class of Horndeski scalar-tensor gravity theory for warm inflation has been analyzed (see Refs.~\cite{HG1, HG2, HG3, HG4} for additional applications). Warm inflation is an interesting scenario where radiation and ultrarelativistic matter survive to the early-time accelerated expansion. As a result, when inflation ends, any reheating mechanism for the particle production is required. 

We have proposed some models to show how warm inflation can take place in the Horndeski framework in the weak and in the strong dissipation regimes. In order to obtain a realistic theory for inflation, the study of cosmological perturbations plays an important role, since the inhomogeneities of our Universe are a direct consequence of perturbations left at the end of the early-time aceleration. Thus, we have investigated the scalar perturbations in the weak dissipation regime and in the strong dissipation regime. In the last case, the perturbations around the Horndeski field are no longer generated by quantum fluctuations, but are given by the thermal interaction with the radiation field. The spectral indexes for our models have been derived and their values have been confronted with the last cosmological data. In the same way, we also analyzed the tensor-to-scalar ratio of tensorial perturbations. In general, Horndeski gravity for warm inflation can bring to a realistic description for the early-time inflation. 

It may be interesting to study the feauture of our model in the presence of the quantum corrections generated at high energy. For example, one may consider a $R^2$-correction to the gravitational action (see Refs.~\cite{R21, R22}). 
Despite to the fact that the presence of the $R^2$-term does not modify the de Sitter solution describing inflation (unless one would not consider an asymptotic solution for which $R^2$-gravity is totally dominant), it may play a fundamental role in the dynamics of perturbations leading to significative corrections of the predictions of our model at least in the weak dissipation regime. On the other hand, in the strong dissipation regime where the perturbations are given by the thermal interactions of the Horndeski field with radiation, the $R^2$-term does not modify the temperature of the de Sitter Universe (see Eq.~(\ref{64})) and we can reasonably argue that our results remain still valid.


\end{document}